\documentclass[aps,pra,twocolumn,showpacs,amsmath,amssymb,groupedaddress]{revtex4-1}
\usepackage{times}
\usepackage{graphicx}
\begin{document}
\title{Thermal-assisted Anisotropy and Anisotropy-driven Instability in the Superfluidity State of Two-species Fermionic Polar Molecules}
\author{Renyuan Liao}\email{rliao08@gmail.com}
\author{Wu-Ming Liu}
\affiliation{National Laboratory for Condensed Matter Physics, Institute of Physics, Chinese Academy of Sciences, Beijing 100190, China}
\date{\today}
\begin{abstract}
 We study the superfluid state of two-species heteronuclear Fermi gases with isotropic contact and anisotropic long-range dipolar interactions. By explicitly taking account of Fock exchange contribution, we derive self-consistent equations describing the pairing states in the system. Exploiting the symmetry of the system, we developed an efficient way of solving the self-consistent equations by exploiting the symmetries. We find that the temperature tends to increase the anisotropy of the pairing state, which is rather counterintuitive. We study the anisotropic properties of the system by examining the angular dependence of the number density distribution, the excitation spectrum and the pair correlation function. The competing effects of the contact interaction and the dipolar interaction upon the anisotropy are revealed. We derive and compute the superfluid mass density $\rho_{ij}$ for the system. Astonishingly, we find that $\rho_{zz}$ becomes negative above some certain temperature $T^*$($T<T_c$), signaling some instability of the system. This suggests that the elusive FFLO state may be observed in experiments, due to an anisotropic state with a spontaneously generated superflow.
\end{abstract}
\maketitle
\section{Introduction}
 The recent experimental realization and coherent control of high phase-space density quantum gas of the polar molecules $^{40}K^{87}Rb$ ~\cite{JIN08,JIN10,JIN102} has stimulated  great interests in studying the effect of  anisotropic long-range dipole-dipole interactions. Furthermore, the large dipole moment of the polar molecules can be tuned using an external electric field; This has a range of applications such as the control of ultracold chemical reactions~\cite{JIN102}, the design of a platform for quantum information processing~\cite{DEM02,AND06} and the realization of novel quantum many-body systems~\cite{MIC06,BAR08}.

 For the fermionic polar molecules, dipolar interactions has led to various new phenomena, including fractional quantum Hall effect~\cite{Bara05}, Wigner crystalization~\cite{BARA08}, nematic phases~\cite{FRAD09}, interlayer superfluidity~\cite{PIK10} and density wave patterns~\cite{MIK11}. Two fundamental properties of dipolar Fermi gases are superfluid pairing~\cite{Bara02,Bara04,BRU08,CON10,ZHA10,LIA10} and Fermi surface deformation~\cite{MIY08,SOG09,ZHA09,BLA10}, originating from the partially attractive nature of the dipolar interaction and anisotropic Fock exchange interaction. For dipolar Fermi gases with two hyperfine states, one can tune not only the dipole-dipole interaction by a fast rotating orienting field~\cite{STE02}, but also the s-wave interspecies interaction via a Feshbach resonance. Therefore, one naturally expects rich physics will emerge as a result of the interplay of the anisotropic long-range dipole interaction and short-range s-wave interaction.

  There have been intense theoretical efforts on the properties of the normal phase at zero temperature, including zero sound propagation~\cite{RON10}, Fermi liquid description~\cite{CHA10}, and expansion dynamics~\cite{GOR03,HE08,SOG09,LIM10}.  At finite temperature, there have been several theoretical efforts. Endo~\cite{END10} et al. developed a variational approach, which allowed them to assess the effect of temperature on the position- and momentum-space distortion. Zhang and Yi~\cite{ZYI10} explored the system deformation and stability by using a Monte Carlo method. Baillie and  Blakie~\cite{BLA10} studied thermodyamics and correlation properties of the system by employing a semiclassical self-consistent Hartree-Fock calculation.

  In this paper, we consider BCS pairing of two-species dipolar Fermi gases at finite temperature, aiming to investigate thermodynamics and anisotropic properties in the superfluid state. To fully assess the effect of Fock exchange term, it is essential to explicitly take account of the Fock exchange term in a self-consistent way. The content of the paper is organized as follows. In Sec. II, we present our model Hamiltonian for a homogeneous two-species dipolar Fermi gas and develop a mean-field description for the superfluid state at finite temperature. In Sec. III, We present theoretical studies and numerical results on the anisotropic properties and the thermodynamics of the system . In Sec. IV, we derive and compute superfluid mass density, then explore the consequences. Finally, we conclude in Sec. V.

\section{Model Hamiltonian and Mean-field Description}
We consider a homogeneous gas of two species of fermionic heteronuclear molecules $\sigma=\uparrow$ and $\downarrow$. For simplicity, we further assume that each species has the same mass, density and dipole moment. The electric dipoles of the molecules with moment $\bold{d}$ are oriented along the $z$ axis by a sufficient strong external electric field such that the spin-independent part of the electronic dipole-dipole interaction becomes $V_{dd}(q)=(4\pi/3)(3\cos^2{\theta_{\bold{q}}-1)}$, with $\theta_\bold{q}$ being the angle between momentum transfer $\bold{q}$ and the direction of z axis in which the dipoles are aligned. In addition, we assume that the molecules also interact via a contact interaction with strength $g$. This system is described by the following Hamiltonian
\begin{eqnarray}
   H-\mu n&=&\sum_{\bold{k}\sigma}(\epsilon_{\bold{k}}-\mu)c_{\bold{k}\sigma}^\dagger c_{\bold{k}\sigma}\nonumber\\
   &+&\frac{1}{2\mathcal{V}}\sum_{\bold{kpq}\sigma\sigma^\prime}V_{\sigma\sigma^\prime}(\bold{q})c_{\bold{k}+\bold{q}\sigma}^\dagger c_{\bold{p}-\bold{q}\sigma^\prime}^\dagger c_{\bold{p}\sigma^\prime}c_{\bold{k}\sigma},
\end{eqnarray}
where $\mu$ is the chemical potential, $n$ is the total number density, $\mathcal{V}$ is the volume, and $\epsilon_{\bold{k}}=k^2/2m$ (where we have set $\hbar=1$). The interaction potential $V_{\sigma\sigma^\prime}(\bold{q})=g\delta_{\sigma,-\sigma^\prime}+V_{dd}(\bold{q})$ contains both dipole-dipole and contact interactions. Anticipating the importance of the Fock exchange term, we decouple the interaction in all three channels~\cite{SIM06}: direct channel, exchange channel and Cooper channel, resulting in the following effective mean-field Hamiltonian
\begin{eqnarray}
   H_{MF}&\!=\!&\sum_{\bold{k}\sigma}\xi_{\bold{k}\sigma}c_{\bold{k}\sigma}^\dagger c_{\bold{k}\sigma}\nonumber\\
   &\!+\!&\frac{1}{2}\sum_{\bold{k}\sigma\sigma^\prime}[\Delta_{\sigma^\prime\sigma}^*(\bold{k})c_{-\bold{k}\sigma^\prime} c_{\bold{k}\sigma}\!+\!\Delta_{\sigma^\prime\sigma}(\bold{k})c_{\bold{k}\sigma}^\dagger c_{-\bold{k}\sigma^\prime}^\dagger].\label{eq:two}
\end{eqnarray}
Here $\xi_{\bold{k}\sigma}=\epsilon_{\bold{k}}-\mu+gn/2+\Sigma_{\bold{k}\sigma}$, with self-consistent mean-fields defined as
\begin{eqnarray}
  \Sigma_{\bold{k}\sigma}&=&-\frac{1}{\mathcal{V}}\sum_{\bold{p}}V_{dd}(\bold{p}-\bold{k})<c_{\bold{p}\sigma}^\dagger c_{\bold{k}\sigma}>,\\
  \Delta_{\sigma^\prime\sigma}(\bold{k})&=&\frac{1}{\mathcal{V}}\sum_{\bold{p}}V_{\sigma\sigma^\prime}(\bold{p}-\bold{k})<c_{-\bold{k}\sigma^\prime}c_{\bold{k}\sigma}>.
\end{eqnarray}
Some comments are in order: The contact interaction affects the single-particle spectrum by shifting the chemical potential, which may be redefined as $\tilde{\mu}=\mu-gn/2$. The self-energy $\Sigma_{\bold{k}\sigma}$ encodes the anisotropic dipolar contribution from Fock exchange term to the dressed single-particle spectrum, which justifies our treatment. In addition, both parts of the interaction contributes to the pairing field.

The Hamiltonian ($\ref{eq:two}$) is diagonalized by invoking the Bogoliubov transformation~\cite{VOL90}. We obtain self-consistent equation for the self-energy, the order parameter and the number density
\begin{eqnarray}
\Sigma_{\bold{k}\sigma}=-\sum_{\bold{p}}V_{dd}(\bold{p}-\bold{k})\left[\frac{1}{2}-\frac{\xi_{\bold{p}}}{2E_{\bold{p}\sigma}}
\tanh{\frac{\beta E_{\bold{p}\sigma}}{2}}\right],\label{eq:sigma}\\
\Delta_{\sigma\sigma^\prime}(\bold{k})=-\sum_{\bold{p}}V_{\sigma\sigma^\prime}(\bold{p}-\bold{k})\frac{\Delta_{\sigma\sigma^\prime}(\bold{p})}{2E_{\bold{p}\sigma}}\tanh{\frac{\beta E_{\bold{p}\sigma}}{2}},\label{eq:gap}\\
n=\sum_{\bold{k}}n_{\bold{k}}=\sum_{\bold{k}\sigma}\frac{1}{2}\left[1-\frac{\xi_{\bold{p}}}{E_{\bold{p}\sigma}}
\tanh{\frac{\beta E_{\bold{p}\sigma}}{2}}\right],\label{eq:number}
\end{eqnarray}
where $E_{\bold{k}\sigma}=\sqrt{\xi_{\bold{k}\sigma}^2+\sum_{\sigma^\prime}|\Delta_{\sigma\sigma^\prime}(\bold{k})|^2}=E_{\bold{k}}$ is the quasiparticle spectrum and $\beta=1/k_BT$ is the inverse temperature. Equation ($\ref{eq:gap}$) formally diverges. For the contact interaction, one can eliminate the interaction strength $g$ in favor of the s-wave scattering length $a_s$ using $1/g=m/(4\pi a_s)-1/\mathcal{V}\sum_{\bold{k}}(2\epsilon_{\bold{k}})^{-1}$. The dipolar interaction  can be regularized by replacing the bare interaction $V_{\sigma\sigma^\prime}(\bold{k}-\bold{p})$ with the vertex function $\Gamma_{\sigma\sigma^\prime}(\bold{k}-\bold{p})$ as explained in Refs.~\cite{Bara02,GUR07}. To first order in Born approximation, the gap equations become
\begin{eqnarray}
  \Delta_{\sigma\sigma^\prime}(\bold{k})=\sum_{\bold{p}}V_{\sigma\sigma^\prime}(\bold{p}-\bold{k})\left[\frac{\tanh{\beta E_{\bold{p}\sigma}}}{2E_{\bold{p}\sigma}}-\frac{1}{2\epsilon_{\bold{p}}}\right].
\end{eqnarray}
The above equation, together with Eqs.~($\ref{eq:sigma}$) and ($\ref{eq:number}$), comprises a complete description of the dipolar Fermi gas and needs to be solve self-consistently.

Due to the symmetry of the interaction potential, the momentum distribution, order parameter and self energy possess azimuthal symmetry in thermal equilibrium, as could be seen from the self-consistent equations by integrating out the azimuthal degrees of freedom $\phi_{\bold{k}}$. Therefore physical quantities such as $n_\bold{k}$, $\Delta_{\sigma\sigma^\prime}(\bold{k})$ and $\Sigma_{\bold{k}\sigma}$ are only functions of $(k,\theta_{\bold{k}})$. Numerically, we parameterized these functions by two dimensional grids living on the domain $\Omega=[0,k_c]\times[0,\pi]$ (where $k_c$ is the momentum cutoff). In this study, we focus on the spin singlet BCS pairing, and suppress the spin index for order parameter by writing $\Delta_{\downarrow\uparrow}(\bold{k})\equiv\Delta(\bold{k})$. The singlet order parameter possesses inversion symmetry as well as azimuthal symmetry $\Delta(k,\theta_{\bold{k}})=\Delta(k,\pi-\theta_{\bold{k}})$. It is easily to verify that $n_\bold{k}$ and $\Sigma_{\bold{k}}$  share the same symmetry. Thus the results presented below is limited to $0\leq\theta_\bold{k}\leq\pi/2$. In our calculation, we parameterize the dipolar interaction by the dimensionless coupling parameter $C_{dd}=md^2(n_\sigma)^{1/3}$, where $n_\sigma=k_F^3/(6\pi^2)=n/2$, and the Fermi energy is $E_F=k_F^2/2m$. To make our mean-field study qualitatively valid and quantitatively reliable, we will restrict ourself to studying weak-coupling BCS regime.

\section{Anisotropic Singlet Superfluid at Finite Temperature}
\begin{figure}[t]
{\scalebox{0.30}{\includegraphics[clip,angle=0]{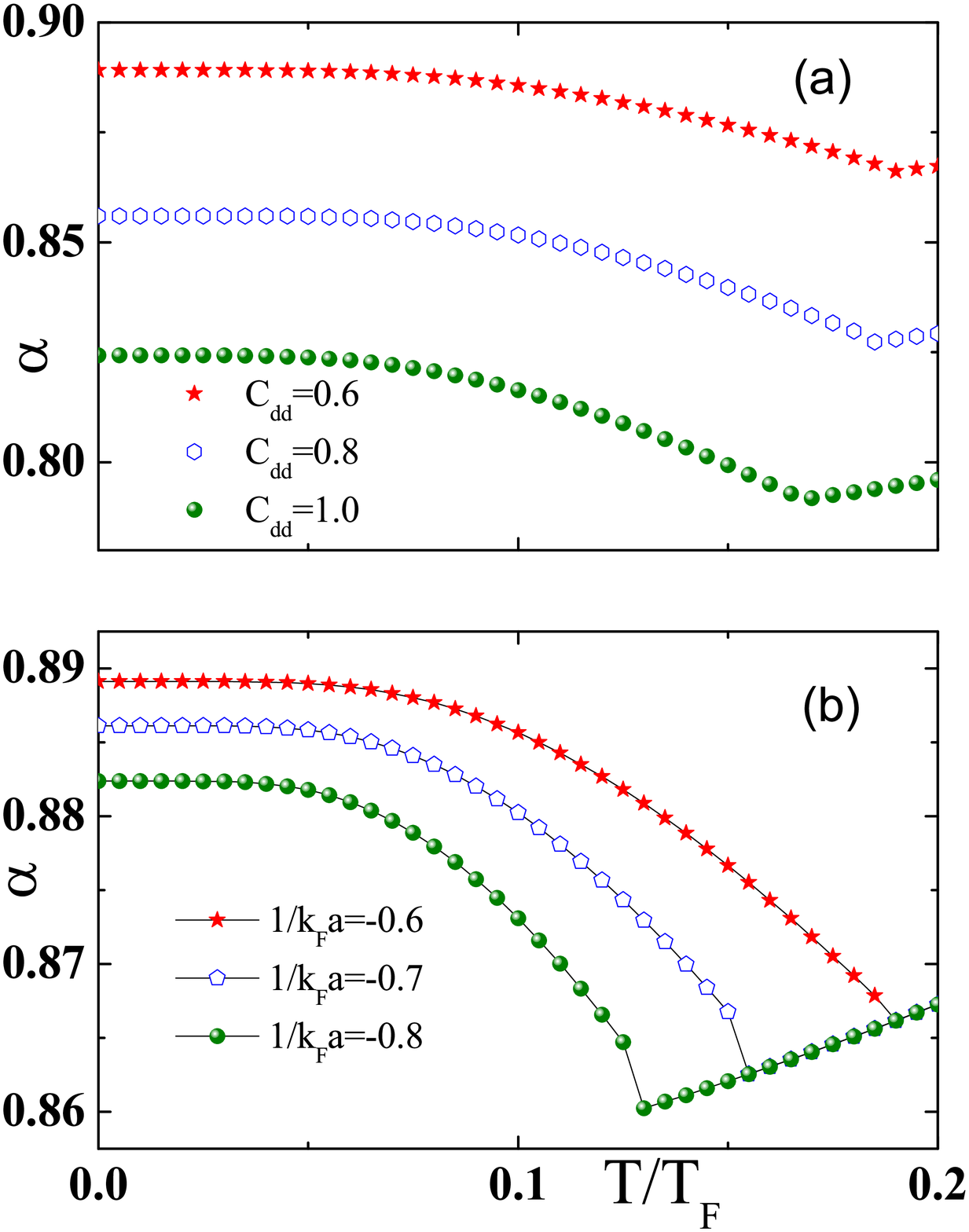}}}
\caption{(Color online) The temperature dependence of the deformation parameter $\alpha$. (a) At s-wave coupling strength $1/k_Fa=-0.6$ and for different dipolar coupling strength $C_{dd}$; (b) At dipolar coupling strength $C_{dd}=0.6$ and for different s-wave coupling strength $1/k_Fa$. The minimum point in each curve corresponds to the position of phase transition between the superfluid state and the normal state. $T_F$ is the Fermi temperature.}
\label{Fig1}
\end{figure}

 In the normal phase at zero temperature, the Fermi surface of the dipolar gas has an ellipsoid shape~\cite{MIY08}: $n_{\bold{k}}=\Theta(k_F^2-\alpha^2k_z^2-k_x^2/\alpha-k_y^2/\alpha)$, where $\alpha$ is the deformation parameter. In a previous study~\cite{LIA10}, we generalized the definition of the deformation parameter $\alpha$ to measure the anisotropy of the single particle momentum distribution with a similar strategy. The angular distribution $n_\theta(\theta_{\bold{k}})$ can be obtained by integrating out the magnitude and azimuthal angle of the momentum  $n_\theta(\theta_{\bold{k}})=\int n(\bold{k})k^2dk/(2\pi^2)$. Since $n_\theta(0)/n_\theta(\pi/2)=k_z^3/k_x^3$ and $\alpha^2k_z^2=k_x^2/\alpha$, we deduce that $\alpha=[n_\theta(\pi/2)/n_\theta(0)]^{2/9}$. In Fig.~\ref{Fig1}, the deformation parameter $\alpha$ is plotted as a function of the temperature. For a fixed for s-wave coupling strength $1/k_Fa=-0.6$, we show, in the panel (a), the deformation parameter $\alpha$ for three typical dipolar coupling strength $C_{dd}=0.6$, $0.8$ and $1$. In each curve, the deformation parameter $\alpha$ develops a minimum at a critical temperature $T_c$ separating the superfluid state from the normal state. The calculated critical temperature $T_c$ is $0.175T_F$, $0.166T_F$ and $0.155T_F$ for $C_{dd}=0.6$, $0.8$ and $1$, respectively. In the normal phase where ($T>T_c$), $\alpha$ increases as the temperature increases. This is expected as increasing the temperature tends to make the system more isotropic by reducing the effects of anisotropic interaction. In the superfluid state, the deformation parameter $\alpha$ decreases as the temperature increases, which is rather counterintuitive. This may be understood in the following way: increasing the temperature tends to weaken the BCS singlet pairing state which favors an isotropic momentum distribution. In both the normal state and the superfluid state, for fixed s-wave coupling, the larger the dipolar coupling strength $C_{dd}$, the smaller the deformation parameter $\alpha$. For a fixed $C_{dd}=0.6$, as shown in the panel (b), the deformation parameter decreases as the temperature increases for all three typical s-wave interaction strength $1/k_Fa=-0.6$, $-0.7$ and $-0.8$. The critical temperature $T_c$ as well as the deformation parameter increases as one increases the s-wave coupling strength. In the normal phase, increasing the s-wave interaction strength has no effect on the deformation parameter, as the pairing channel closes and the effective chemical potential remains the same so as to fixed the average particle number.

To further reveal the anisotropic nature of the pairing states, we investigate the pair correlation function and the quasiparticle excitation spectrum. The pair correlation function can be evaluated as~\cite{PAR07}
\begin{eqnarray*}
  C_{\uparrow\downarrow}(\bold{k}_1,\bold{k}_2)&=&<n_{\bold{k}_1\uparrow}n_{\bold{k}_2\downarrow}>-<n_{\bold{k}_1\uparrow}><n_{\bold{k}_2\downarrow}>\\
  &=&\delta_{\bold{k}_1,-\bold{k}_2}\frac{|\Delta_{\bold{k}}|^2}{4E_\bold{k}^2}\tanh^2{\frac{\beta E_\bold{k}}{2}}.
\end{eqnarray*}
\begin{figure}[t]
{\scalebox{0.30}{\includegraphics[clip,angle=0]{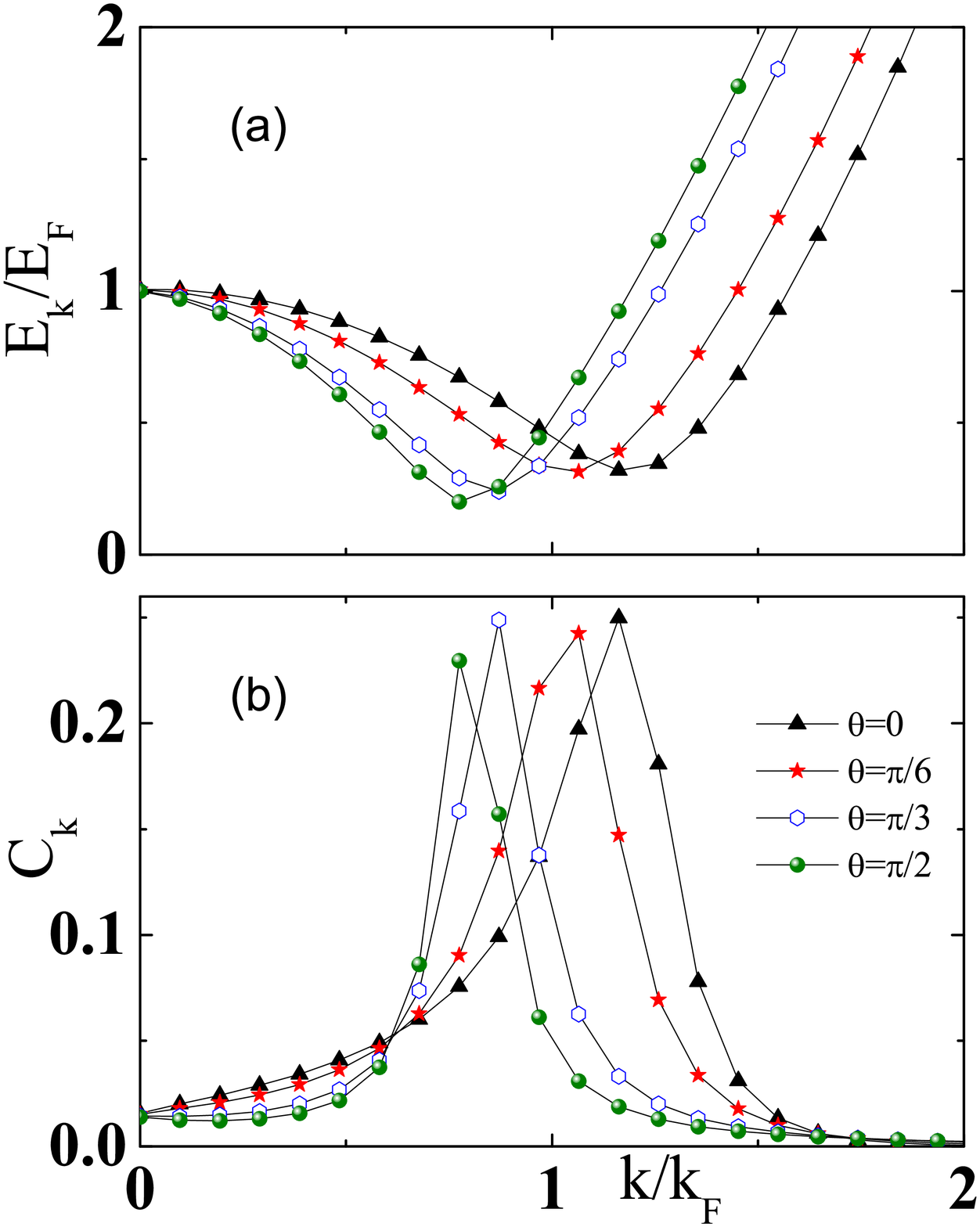}}}
\caption{(Color online) (a) The quasiparticle excitation spectrum $E_\bold{k}$ and (b) the pair correlation function $C_\bold{k}$ along various polar angles as a function of momentum magnitude for the s-wave coupling strength $1/k_Fa=-0.6$ and the dipolar coupling strength $C_{dd}=1$ at zero temperature. }
\label{Fig2}
\end{figure}
As shown in Fig.~\ref{Fig2}a, the quasiparticle excitation spectrum $E_\bold{k}$ starts at the same value from zero momentum and evolves anisotropically in momentum space. Along any polar angle, $E_\bold{k}$ decreases at small momentum magnitude and increases at high momentum magnitude; it develops a minimum which is greater than zero, signaling that it is a gapped superfluid. The locus of the minimum varies from a small momentum magnitude to a large momentum magnitude as the polar angle $\theta_\bold{k}$ decreases from $\pi/2$ to $0$, indicating a distortion along the $z$ axis. The pair correlation function $C_\bold{k}$ is shown in Fig.~\ref{Fig2}b. It is interesting to notice that for a fixed polar angle, there exists some typical momentum magnitude at which the pair correlation function $C_\bold{k}$ achieves a maximum. Remarkably, at this typical momentum, the quasiparticle excitation spectrum $E_\bold{k}$ develops a minimum. As the polar angle changes from $0$ to $\pi/2$, the momentum magnitude at which the maximum pairing occurs decreases. At a large momentum magnitude, the pair correlation decays rapidly with $1/k^4$ asymptotically.

\begin{figure}[t]
{\scalebox{0.30}{\includegraphics[clip,angle=0]{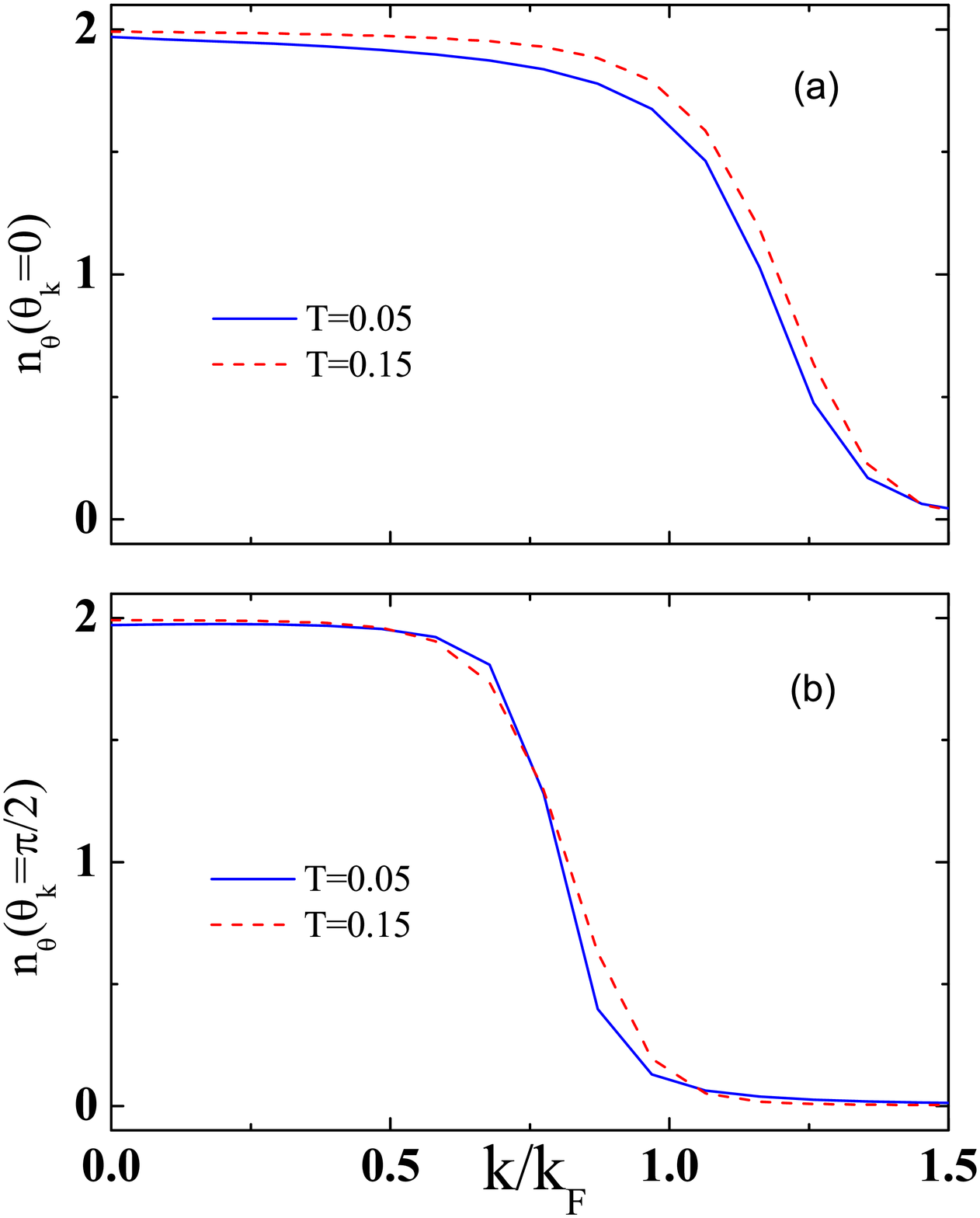}}}
\caption{(Color online) The radial number occupancy $n_\theta(\theta_{\bold{k}})$ at s-wave coupling strength $1/k_Fa=-0.6$ and dipolar coupling strength $C_{dd}=1.0$ for (a) the polar direction $\theta_{\bold{k}}=0$ and (b) the axial direction $\theta_{\bold{k}}=\pi/2$. The solid blue curves are for $T/T_F=0.05$, and the dash red curves are for $T/T_F=0.15$. $T_F=E_F/k_B$ is the Fermi temperature.}
\label{Fig3}
\end{figure}

To assess the temperature effect on the anisotropic properties of the system, we study momentum space distribution of the number density. The plot in panel (a) and (b) of Fig.~\ref{Fig3} corresponds to radial momentum distribution of the number density along the polar ($\theta_\bold{k}=0$) and axial direction ($\theta_{\bold{k}}=\pi/2$), respectively. At $\theta_{\bold{k}}=0$, the number occupation at any given momentum magnitude increases slightly as the temperature is raised from $T/T_F=0.05$ to $T/T_F=0.15$. In contrast to the polar direction, the situation at $\theta_{\bold{k}}=\pi/2$ looks quite different: for $k<k_F$, the number occupancy for $T/T_F=0.05$ and $T/T_F=0.15$ coincides at two typical momentum magnitudes; For momentum magnitude lies in the range of $0.5k_F<k<0.6k_F$, the number occupancy for lower temperature exceeds that of higher temperature; the trend appears to be reversed for $0.6k_F<k<1.0k_F$. Judging from the cases of $\theta_\bold{k}=0$ and $\theta_\bold{k}=0$, one can largely deduce that as the temperature increases, the number occupancy transfers from the axial direction to the polar direction, rendering the system more anisotropic. This is consistent with the temperature dependence of the deformation parameter $\alpha$ shown in Fig.~\ref{Fig1}.

\begin{figure}[t]
{\scalebox{0.30}{\includegraphics[clip,angle=0]{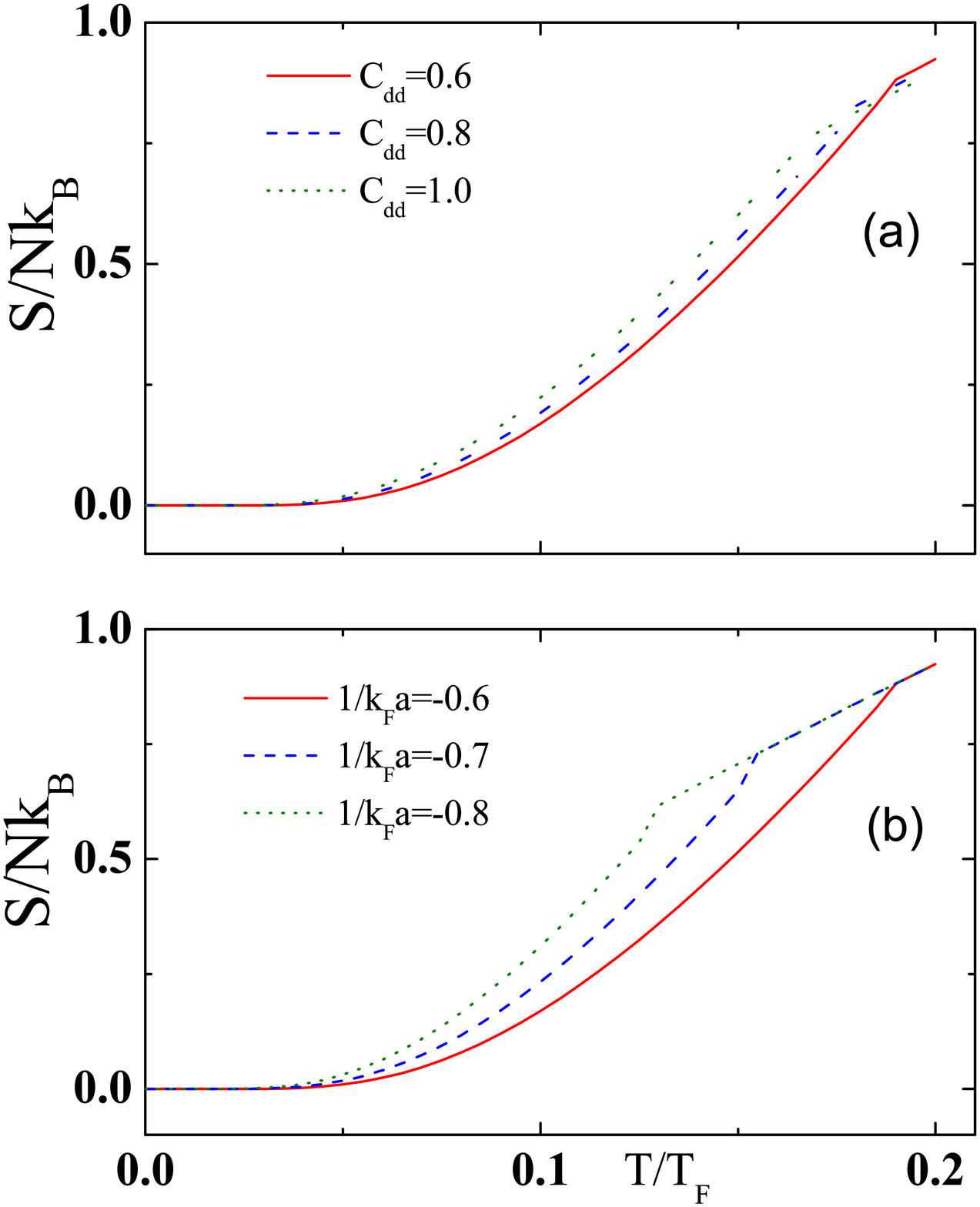}}}
\caption{(Color online) The entropy $S/(Nk_B)$ as a function of the temperature. (a) At s-wave coupling strength $1/k_Fa=-0.6$ and for different dipolar coupling strength $C_{dd}$; (b) At dipolar coupling strength $C_{dd}=0.6$ and for different s-wave coupling strength $1/k_Fa$. The linear (high temperature) part of the curves corresponds to the normal phase, characteristic of Fermi liquid behavior.}
\label{Fig4}
\end{figure}

The Entropy provides useful insights into the thermodynamics of the superfluid pairing states. At mean-field level, the system may be regarded as a collection of noninteracting quasiparticles with dispersion $E_{\bold{k}\sigma}$, obeying Fermi statistics. Thus, the entropy of the system is evaluated as
\begin{eqnarray}
   S&=&-k_B\sum_{\bold{k}\sigma}\left[n_F(E_{\bold{k}\sigma})\ln{n_F(E_{\bold{k}\sigma})}\right]\nonumber\\
                        & &-k_B\sum_{\bold{k}\sigma}\left[(1-n_F(E_{\bold{k}\sigma}))\ln{(1-n_F(E_{\bold{k}\sigma}))}\right]\nonumber\\
                        &=&k_B\sum_{\bold{k}\sigma}\left[\beta E_{\bold{k}\sigma}n_F(E_{\bold{k}\sigma})+\ln(1+e^{-\beta E_{\bold{k}\sigma}})\right],
\end{eqnarray}
where $n_F$ is the Fermi function defined  by $n_F(E_{\bold{k}\sigma})=1/[1+\exp{(\beta E_{\bold{k}\sigma})}]$.
In Fig.~\ref{Fig4}a, the entropy is plotted as a function of the temperature for fixed s-wave coupling strength $1/k_Fa=-0.6$ and for different dipolar coupling strength.
As seen in the figure, for every dipolar coupling strength $C_{dd}$ there exists a critical temperature $T_c$ separating the normal phase from the ordered pairing phase. The entropy in the normal phase has characteristic linear temperature dependence, a signature of the Fermi liquid behavior. We find that, in the superfluid phase, a larger dipolar interaction strength leads to a higher entropy, which is opposite to what is found in a polarized normal phase~\cite{BLA10,ZYI10}. As a measure of the degree of disorder, a higher entropy characterizes a less ordered state. This finding is consistent with the established argument~\cite{LIA10} that in the singlet pairing phase the anisotropic dipolar interaction is a competing effect as opposed to the s-wave contact interaction. In Fig.~\ref{Fig4}b, the entropy decreases as the s-wave coupling strength $1/k_Fa$ increases. When the system is in the normal phase, increasing the s-wave interaction has no effect on the entropy. This is due to the fact that, in the normal state the pairing channel is closed, and the effective chemical potential $\tilde{\mu}=\mu-gn/2$ is the same so as to fix the average particle number.
\section{Superfluid density}
The superfluid density is a fundamental quantity describing the response to a rotation as well as in two-fluid collisional hydrodynamics. In the following, we will obtain the superfluid density using the thermodynamic potential in the presence of a superfluid flow. To impose a current, one applies a ``phase twist" to the order parameter $\Delta(x)$: $\Delta(x)\rightarrow \Delta(x)e^{i\bold{Q}\cdot\bold{r}}$. The superfluid velocity $\bold{v_s}$ associated with this imposed phase twist is $\bold{v_s}=\bold{Q}/M$, and $M$ is the Cooper pair mass $M=2m$. The grand potential with superfluid velocity $\bold{v_s}$ could be written as~\cite{ZHU06,GRI06}
\begin{eqnarray}
 \Omega(\bold{v_s})&=&\Omega(0)+\bold{j_s}\cdot\bold{v_s}+\frac{1}{2}\rho_{ij}(\bold{v_s})_i(\bold{v_s})_j+\cdot \cdot\cdot,\label{eq:omega1}\\
  \Omega(\bold{v_s})&-&\Omega(0)=-\frac{1}{\beta}(Tr\ln{G_{s}^{-1}}-Tr\ln{G^{-1}})\label{eq:omega2}.
\end{eqnarray}

In the above, $\bold{j_s}$ is the superfluid current, $\rho_{ij}$ is the superfluid (mass) density tensor, the trace operator $Tr$ runs over both spin space and space-time coordinate $x=(\bold{r},\tau)$, and $G^{-1}(x,x^\prime)$ is the inverse of the Green's function in the Nambu space
\begin{eqnarray}
    &&G^{-1}(x,x^\prime)=\delta(x-x^\prime)\times\nonumber\\
    &&    \begin{pmatrix}
                -\partial_\tau-\frac{\hat{\bold{P}^2}}{2m}+\mu-\Sigma(x) &   \Delta(x)\\
                \Delta^*(x) & -\partial_\tau+\frac{\hat{\bold{P}^2}}{2m}-\mu+\Sigma(x)
           \end{pmatrix}\label{eq:Green}
\end{eqnarray}
\begin{figure}[t]
{\scalebox{0.30}{\includegraphics[clip,angle=0]{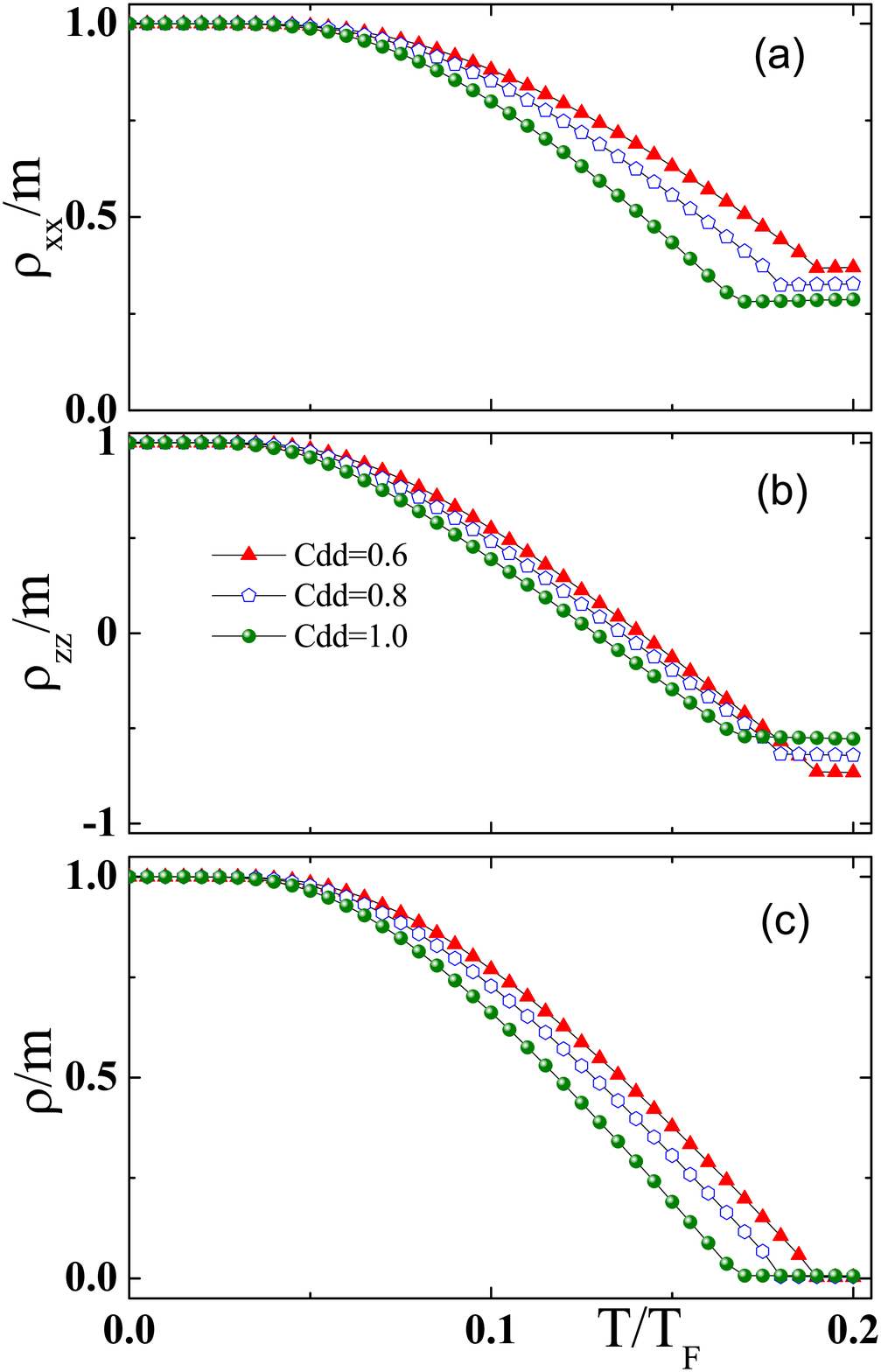}}}
\caption{(Color online) The superfluid density tensor $\rho_{ij}$ as a function of temperature for s-wave coupling strength $1/k_Fa=-0.6$ and for various dipolar coupling strength $C_{dd}$: (a) $\rho_{xx}$, (b) $\rho_{zz}$ and (c) $\rho=\left(\rho_{xx}+\rho_{yy}+\rho_{zz}\right)/3.$}
\label{Fig5}
\end{figure}
To bring $Tr\ln{G_{s}^{-1}}$ into a more convenient form, we introduce a unitary matrix
\begin{eqnarray}
  U=\begin{pmatrix}
        e^{i\bold{Q}.\bold{r}/2} &  0 \\
        0   &   e^{-i\bold{Q}\cdot\bold{r}/2}
    \end{pmatrix}.
\end{eqnarray}
Employing $Tr\ln{G_{s}^{-1}}=Tr\ln{\left[U^\dagger G_s^{-1}U\right]}=Tr\ln{\tilde{G}^{-1}}$, we obtain
\begin{eqnarray}
   &&\tilde{G}^{-1}(x,x^\prime)=\delta(x-x^\prime)\times\nonumber\\
   &&\begin{pmatrix}
                      -\partial_\tau-\frac{(\hat{\bold{P}}+\frac{\bold{Q}}{2})^2}{2m}+\mu-\Sigma(x) &   \Delta(x)\\
                      \Delta^*(x)   & -\partial_\tau+\frac{(\hat{\bold{P}}-\frac{\bold{Q}}{2})^2}{2m}-\mu+\Sigma(x)
                      \end{pmatrix}\nonumber\\. \label{eq:Greens}
\end{eqnarray}
Transforming Eqs.~($\ref{eq:Green}$) and ~($\ref{eq:Greens}$) to the momentum-frequency representation, we obtain
\begin{eqnarray}
  \tilde{G}^{-1}(\bold{k},iw_n)=G^{-1}-\frac{\bold{k}\cdot\bold{Q}}{2m}\otimes I-\frac{\bold{Q}^2}{8m}\otimes\sigma_z.
\end{eqnarray}
Therefore, one can perform the derivative expansion,
\begin{eqnarray}
   Tr\ln{G^{-1}}-Tr\ln{\tilde{G}^{-1}}&=&\bold{k}\cdot\bold{v_s}TrG+\frac{m\bold{v_s}^2}{2} Tr(G\sigma_3)\nonumber\\
   &+&\frac{\left(\bold{k}\cdot\bold{v_s}\right)^2}{2}Tr(GG)+\cdot\cdot\cdot. \label{eq:omega3}
\end{eqnarray}
Combining Eqs.~($\ref{eq:omega1}$), ($\ref{eq:omega2}$) and ($\ref{eq:omega3}$), we readily obtain the expression for superfluid density tensor
\begin{eqnarray}
   \rho_{ij}&=&\frac{\partial^2\Omega}{\partial(v_s)_i\partial(v_s)_j}|_{v_s\rightarrow 0}\nonumber\\
            &=&\frac{\delta_{ij}}{\beta}\left[mTr(G\sigma_3)+k_i^2Tr(GG)\right]\nonumber\\
            &=&\delta_{ij}\left[mn+\sum_{\bold{k},iw_n}\frac{k_i^2}{\beta}\left(G_{11}^2+G_{22}^2+2G_{12}G_{21}\right)\right].\label{eq:rho}
\end{eqnarray}
To proceed further, one needs the Green's function $G(\bold{k},iw_n)$ obtained from Eq.~($\ref{eq:Green}$)
\begin{eqnarray}
    G(\bold{k},iw_n)= \frac{1}{(iw_n)^2-E_\bold{k}^2}
                     \begin{pmatrix}
                     iw_n+\xi_\bold{k} & -\Delta_\bold{k}^* \\ -\Delta_\bold{k} & iw_n-\xi_\bold{k}
                     \end{pmatrix}.\label{eq:green2}
\end{eqnarray}
Substituting Eq.~($\ref{eq:green2}$) into Eq.~($\ref{eq:rho}$) and carrying out the Matsubara frequencies summation, we finally reach
\begin{eqnarray}
\rho_{ij}=\delta_{ij}\left[mn+\sum_\bold{k}2k_i^2n_F^\prime(E_\bold{k})\right].
\end{eqnarray}
The cylindrical symmetry implies that $\rho_{xx}=\rho_{yy}$. The average superfluid density is defined $\rho=(\rho_{xx}+\rho_{yy}+\rho_{zz})/3=(2\rho_{xx}+\rho_{zz})/3$.

The temperature dependence of superfluid density is shown in Fig.~\ref{Fig5} for fixed s-wave interaction strength $1/k_Fa=-0.6$ and for different dipolar interaction strength $C_{dd}=0.6, 0.8$ and $1$. For a given $C_{dd}$, $\rho_{xx}$, $\rho_{yy}$ and $\rho$ decreases as the temperature increases until the temperature reaches the transition temperature $T_c$. Higher $C_{dd}$ weakens the pairing states, leading to lower $\rho_{xx}$, $\rho_{zz}$ and $\rho$. At zero temperature, all the particles contribute to the average superfluid density $\rho$. As seen in panel (c), as the temperature goes up, $\rho$ drops down, and reaches zero at the normal phase. This behavior is similar to what is expected from the picture of the two-fluid model. By close inspection of panel (b), one finds a peculiar behavior: for a given $C_{dd}$, the $\rho_{zz}$ is negative when the temperature is higher than some typical temperature $T^*$ ($T^*<T_C)$, which signals some instability of the system. The FFLO pairing state with a nonzero center of mass momentum along the polar direction is energetically favored due to the negative $\rho_{zz}$. The search for the FFLO state has lasted for more than four decades in many branches of physics. The observation of the FFLO phase in solids proved very difficult and was only achieved recently in the heavy fermion superconductor CeCoIn$_5$~\cite{RAD03}. In cold atom physics, very recently, evidence of the FFLO superfluid has been found~\cite{LIAO10} in the experiment on a one-dimensional Fermi gas.

\section{Conclusion}
To sum up, we have presented a detailed theoretical and numerical study on the singlet superfluid pairing in a two-species dipolar Fermi system at finite temperature. Fascinating consequences of long-range anisotropic dipolar interaction are explored. The competing effects of isotropic s-wave contact interaction and anisotropic dipolar interaction are revealed. We also study finite temperature effect on the system through various physical quantities.
Importantly, We find that increasing the  temperature tends to make the system more anisotropic. In particular, we derive and compute the superfluid density tensor for the system. We argue that the elusive FFLO state may be realized in polar molecules.  The phase space where FFLO state is stable and the properties of the FFLO state deserves further investigation.

The anisotropic properties of momentum distribution and excitation spectrum bear consequences for experimental observations. The single-particle momentum distribution is routinely observed by time-of-flight measurements~\cite{REG05}. Anisotropic features of the quasiparticle spectrum could be probed by radio frequency spectroscopy~\cite{CHI04,GAE10}. The recent experimental work~\cite{LIAO10} paves the way to direct observation and characterization of FFLO pairing. With the experimental progress toward achieving degenerate polar molecules, one may expect that the predictions could be verified experimentally in near future.

\section*{Acknowledgement}
R. Liao acknowledges helpful discussions with J. Brand, Y. Yu and H. Hu. This work was supported by NSFC under grants Nos. $10874235$, $10934010$, $60978019$, the NKBRSFC under grants Nos. $2009$CB$930701$, $2010$CB$922904$, and $2011$CB$921502$, and NSFC-RGC under grants Nos. $11061160490$ and $1386$-N-HKU$748/10$.


\begin{thebibliography}{10}%
\makeatletter
\providecommand \@ifxundefined [1]{%
 \ifx #1\undefined \expandafter \@firstoftwo
 \else \expandafter \@secondoftwo
\fi
}%
\providecommand \@ifnum [1]{%
 \ifnum #1\expandafter \@firstoftwo
 \else \expandafter \@secondoftwo
\fi
}%
\providecommand \enquote [1]{``#1''}%
\providecommand \bibnamefont  [1]{#1}%
\providecommand \bibfnamefont [1]{#1}%
\providecommand \citenamefont [1]{#1}%
\providecommand\href[0]{\@sanitize\@href}%
\providecommand\@href[1]{\endgroup\@@startlink{#1}\endgroup\@@href}%
\providecommand\@@href[1]{#1\@@endlink}%
\providecommand \@sanitize [0]{\begingroup\catcode`\&12\catcode`\#12\relax}%
\@ifxundefined \pdfoutput {\@firstoftwo}{%
 \@ifnum{\z@=\pdfoutput}{\@firstoftwo}{\@secondoftwo}%
}{%
 \providecommand\@@startlink[1]{\leavevmode}%
 \providecommand\@@endlink[0]{}%
}{%
 \providecommand\@@startlink[1]{%
  \leavevmode
  \pdfstartlink
   attr{/Border[0 0 1 ]/H/I/C[0 1 1]}%
   user{/Subtype/Link/A<</Type/Action/S/URI/URI(#1)>>}%
  \relax
 }%
 \providecommand\@@endlink[0]{\pdfendlink}%
}%
\providecommand \url  [0]{\begingroup\@sanitize \@url }%
\providecommand \@url [1]{\endgroup\@href {#1}{\urlprefix}}%
\providecommand \urlprefix [0]{URL }%
\providecommand \Eprint[0]{\href }%
\@ifxundefined \urlstyle {%
  \providecommand \doi [1]{doi:\discretionary{}{}{}#1}%
}{%
  \providecommand \doi [0]{doi:\discretionary{}{}{}\begingroup
  \urlstyle{rm}\Url }%
}%
\providecommand \doibase [0]{http://dx.doi.org/}%
\providecommand \Doi[1]{\href{\doibase#1}}%
\providecommand \bibAnnote [3]{%
  \BibitemShut{#1}%
  \begin{quotation}\noindent
    \textsc{Key:}\ #2\\\textsc{Annotation:}\ #3%
  \end{quotation}%
}%
\providecommand \bibAnnoteFile [2]{%
  \IfFileExists{#2}{\bibAnnote {#1} {#2} {\input{#2}}}{}%
}%
\providecommand \typeout [0]{\immediate \write \m@ne }%
\providecommand \selectlanguage [0]{\@gobble}%
\providecommand \bibinfo [0]{\@secondoftwo}%
\providecommand \bibfield [0]{\@secondoftwo}%
\providecommand \translation [1]{[#1]}%
\providecommand \BibitemOpen[0]{}%
\providecommand \bibitemStop [0]{}%
\providecommand \bibitemNoStop [0]{.\EOS\space}%
\providecommand \EOS [0]{\spacefactor3000\relax}%
\providecommand \BibitemShut [1]{\csname bibitem#1\endcsname}%
%</preamble>
\bibitem{JIN08}%
  \BibitemOpen
  \bibfield{author}{%
  \bibinfo {author} {\bibfnamefont{S.}~\bibnamefont{Ospelkaus}}, \bibinfo
  {author} {\bibfnamefont{A.}~\bibnamefont{P\'{e}er}}, \bibinfo {author}
  {\bibfnamefont{K.-K.}\ \bibnamefont{Ni}}, \bibinfo {author}
  {\bibfnamefont{J.~J.}\ \bibnamefont{Zirbel}}, \bibinfo {author}
  {\bibfnamefont{B.}~\bibnamefont{Neyenhuis}}, \bibinfo {author}
  {\bibfnamefont{S.}~\bibnamefont{Kotochigova}}, \bibinfo {author}
  {\bibfnamefont{P.~S.}\ \bibnamefont{Julienne}}, \bibinfo {author}
  {\bibfnamefont{J.}~\bibnamefont{Ye}},\ and\ \bibinfo {author}
  {\bibfnamefont{D.~S.}\ \bibnamefont{Jin}},\ }%
  \bibfield{journal}{%
  \bibinfo {journal} {Nat. Phys.}\ }%
  \textbf{\bibinfo {volume} {4}},\ \bibinfo {pages} {622} (\bibinfo {year}
  {2008})%
  \bibAnnoteFile{NoStop}{JIN08}%
\bibitem{JIN10}%
  \BibitemOpen
  \bibfield{author}{%
  \bibinfo {author} {\bibfnamefont{S.}~\bibnamefont{Ospelkaus}}, \bibinfo
  {author} {\bibfnamefont{K.-K.}\ \bibnamefont{Ni}}, \bibinfo {author}
  {\bibfnamefont{G.}~\bibnamefont{Qu\'{e}m\'{e}ner}}, \bibinfo {author}
  {\bibfnamefont{B.}~\bibnamefont{Neyenhuis}}, \bibinfo {author}
  {\bibfnamefont{D.}~\bibnamefont{Wang}}, \bibinfo {author}
  {\bibfnamefont{M.~H.~G.}\ \bibnamefont{de~Miranda}}, \bibinfo {author}
  {\bibfnamefont{J.~L.}\ \bibnamefont{Bohn}}, \bibinfo {author}
  {\bibfnamefont{J.}~\bibnamefont{Ye}},\ and\ \bibinfo {author}
  {\bibfnamefont{D.~S.}\ \bibnamefont{Jin}},\ }%
  \bibfield{journal}{%
  \bibinfo {journal} {Phys. Rev. Lett.}\ }%
  \textbf{\bibinfo {volume} {104}},\ \bibinfo {pages} {030402} (\bibinfo {year}
  {2010})%
  \bibAnnoteFile{NoStop}{JIN10}%
\bibitem{JIN102}%
  \BibitemOpen
  \bibfield{author}{%
  \bibinfo {author} {\bibfnamefont{S.}~\bibnamefont{Ospelkaus}}, \bibinfo
  {author} {\bibfnamefont{K.-K.}\ \bibnamefont{Ni}}, \bibinfo {author}
  {\bibfnamefont{D.}~\bibnamefont{Wang}}, \bibinfo {author}
  {\bibfnamefont{M.~H.~G.}\ \bibnamefont{de~Miranda}}, \bibinfo {author}
  {\bibfnamefont{B.}~\bibnamefont{Neyenhuis}}, \bibinfo {author}
  {\bibfnamefont{G.}~\bibnamefont{Qu\'{e}m\'{e}ner}}, \bibinfo {author}
  {\bibfnamefont{P.~S.}\ \bibnamefont{Julienne}}, \bibinfo {author}
  {\bibfnamefont{J.~L.}\ \bibnamefont{Bohn}}, \bibinfo {author}
  {\bibfnamefont{D.~S.}\ \bibnamefont{Jin}},\ and\ \bibinfo {author}
  {\bibfnamefont{J.}~\bibnamefont{Ye}},\ }%
  \bibfield{journal}{%
  \bibinfo {journal} {Science}\ }%
  \textbf{\bibinfo {volume} {327}},\ \bibinfo {pages} {853} (\bibinfo {year}
  {2010})%
  \bibAnnoteFile{NoStop}{JIN102}%
\bibitem{DEM02}%
  \BibitemOpen
  \bibfield{author}{%
  \bibinfo {author} {\bibfnamefont{D.}~\bibnamefont{DeMille}},\ }%
  \bibfield{journal}{%
  \bibinfo {journal} {Phys. Rev. Lett.}\ }%
  \textbf{\bibinfo {volume} {88}},\ \bibinfo {pages} {067901} (\bibinfo {year}
  {2002})%
  \bibAnnoteFile{NoStop}{DEM02}%
\bibitem{AND06}%
  \BibitemOpen
  \bibfield{author}{%
  \bibinfo {author} {\bibfnamefont{A.}~\bibnamefont{Andr\'{e}}}, \bibinfo
  {author} {\bibfnamefont{D.}~\bibnamefont{Demille}}, \bibinfo {author}
  {\bibfnamefont{J.~M.}\ \bibnamefont{Doyle}}, \bibinfo {author}
  {\bibfnamefont{M.~D.}\ \bibnamefont{Lukin}}, \bibinfo {author}
  {\bibfnamefont{S.~E.}\ \bibnamefont{Maxwell}}, \bibinfo {author}
  {\bibfnamefont{P.}~\bibnamefont{Rabl}}, \bibinfo {author}
  {\bibfnamefont{R.~J.}\ \bibnamefont{Schoelkopf}},\ and\ \bibinfo {author}
  {\bibfnamefont{P.}~\bibnamefont{Zoller}},\ }%
  \bibfield{journal}{%
  \bibinfo {journal} {Nat. Phys.}\ }%
  \textbf{\bibinfo {volume} {2}},\ \bibinfo {pages} {636} (\bibinfo {year}
  {2006})%
  \bibAnnoteFile{NoStop}{AND06}%
\bibitem{MIC06}%
  \BibitemOpen
  \bibfield{author}{%
  \bibinfo {author} {\bibfnamefont{A.}~\bibnamefont{Micheli}}, \bibinfo
  {author} {\bibfnamefont{G.~K.}\ \bibnamefont{Brennen}},\ and\ \bibinfo
  {author} {\bibfnamefont{P.}~\bibnamefont{Zoller}},\ }%
  \bibfield{journal}{%
  \bibinfo {journal} {Nat. Phys.}\ }%
  \textbf{\bibinfo {volume} {2}},\ \bibinfo {pages} {341} (\bibinfo {year}
  {2006})%
  \bibAnnoteFile{NoStop}{MIC06}%
\bibitem{BAR08}%
  \BibitemOpen
  \bibfield{author}{%
  \bibinfo {author} {\bibfnamefont{M.}~\bibnamefont{Baranov}},\ }%
  \bibfield{journal}{%
  \bibinfo {journal} {Phys. Rep.}\ }%
  \textbf{\bibinfo {volume} {464}},\ \bibinfo {pages} {71} (\bibinfo {year}
  {2008})%
  \bibAnnoteFile{NoStop}{BAR08}%
\bibitem{Bara05}%
  \BibitemOpen
  \bibfield{author}{%
  \bibinfo {author} {\bibfnamefont{M.~A.}\ \bibnamefont{Baranov}}, \bibinfo
  {author} {\bibfnamefont{K.}~\bibnamefont{Osterloh}},\ and\ \bibinfo {author}
  {\bibfnamefont{M.}~\bibnamefont{Lewenstein}},\ }%
  \bibfield{journal}{%
  \bibinfo {journal} {Phys. Rev. Lett.}\ }%
  \textbf{\bibinfo {volume} {94}},\ \bibinfo {pages} {070404} (\bibinfo {year}
  {2005})%
  \bibAnnoteFile{NoStop}{Bara05}%
\bibitem{BARA08}%
  \BibitemOpen
  \bibfield{author}{%
  \bibinfo {author} {\bibfnamefont{M.~A.}\ \bibnamefont{Baranov}}, \bibinfo
  {author} {\bibfnamefont{H.}~\bibnamefont{Fehrmann}},\ and\ \bibinfo {author}
  {\bibfnamefont{M.}~\bibnamefont{Lewenstein}},\ }%
  \bibfield{journal}{%
  \bibinfo {journal} {Phys. Rev. Lett.}\ }%
  \textbf{\bibinfo {volume} {100}},\ \bibinfo {pages} {200402} (\bibinfo {year}
  {2008})%
  \bibAnnoteFile{NoStop}{BARA08}%
\bibitem{FRAD09}%
  \BibitemOpen
  \bibfield{author}{%
  \bibinfo {author} {\bibfnamefont{B.~M.}\ \bibnamefont{Fregoso}}\ and\
  \bibinfo {author} {\bibfnamefont{E.}~\bibnamefont{Fradkin}},\ }%
  \bibfield{journal}{%
  \bibinfo {journal} {Phys. Rev. Lett.}\ }%
  \textbf{\bibinfo {volume} {103}},\ \bibinfo {pages} {205301} (\bibinfo {year}
  {2009})%
  \bibAnnoteFile{NoStop}{FRAD09}%
\bibitem{PIK10}%
  \BibitemOpen
  \bibfield{author}{%
  \bibinfo {author} {\bibfnamefont{A.}~\bibnamefont{Pikovski}}, \bibinfo
  {author} {\bibfnamefont{M.}~\bibnamefont{Klawunn}}, \bibinfo {author}
  {\bibfnamefont{G.~V.}\ \bibnamefont{Shlyapnikov}},\ and\ \bibinfo {author}
  {\bibfnamefont{L.}~\bibnamefont{Santos}},\ }%
  \bibfield{journal}{%
  \bibinfo {journal} {Phys. Rev. Lett.}\ }%
  \textbf{\bibinfo {volume} {105}},\ \bibinfo {pages} {215302} (\bibinfo {year}
  {2010})%
  \bibAnnoteFile{NoStop}{PIK10}%
\bibitem{MIK11}%
  \BibitemOpen
  \bibfield{author}{%
  \bibinfo {author} {\bibfnamefont{K.}~\bibnamefont{Mikelsons}}\ and\ \bibinfo
  {author} {\bibfnamefont{J.~K.}\ \bibnamefont{Freericks}},\ }%
  \bibfield{journal}{%
  \bibinfo {journal} {Phys. Rev. A}\ }%
  \textbf{\bibinfo {volume} {83}},\ \bibinfo {pages} {043609} (\bibinfo {year}
  {2011})%
  \bibAnnoteFile{NoStop}{MIK11}%
\bibitem{Bara02}%
  \BibitemOpen
  \bibfield{author}{%
  \bibinfo {author} {\bibfnamefont{M.~A.}\ \bibnamefont{Baranov}}, \bibinfo
  {author} {\bibfnamefont{M.~S.}\ \bibnamefont{Ma\`{r}enko}}, \bibinfo {author}
  {\bibfnamefont{V.~.~S.}~\bibnamefont{Rychkov}},\ and\ \bibinfo {author}
  {\bibfnamefont{G.~V.}\ \bibnamefont{Shlyapnikov}},\ }%
  \bibfield{journal}{%
  \bibinfo {journal} {Phys. Rev. A}\ }%
  \textbf{\bibinfo {volume} {66}},\ \bibinfo {pages} {013606} (\bibinfo {year}
  {2002})%
  \bibAnnoteFile{NoStop}{Bara02}%
\bibitem{Bara04}%
  \BibitemOpen
  \bibfield{author}{%
  \bibinfo {author} {\bibfnamefont{M.~A.}\ \bibnamefont{Baranov}}, \bibinfo
  {author} {\bibfnamefont{L.}~\bibnamefont{Dobrek}},\ and\ \bibinfo {author}
  {\bibfnamefont{M.}~\bibnamefont{Lewenstein}},\ }%
  \bibfield{journal}{%
  \bibinfo {journal} {Phys. Rev. Lett.}\ }%
  \textbf{\bibinfo {volume} {92}},\ \bibinfo {pages} {250403} (\bibinfo {year}
  {2004})%
  \bibAnnoteFile{NoStop}{Bara04}%
\bibitem{BRU08}%
  \BibitemOpen
  \bibfield{author}{%
  \bibinfo {author} {\bibfnamefont{G.~M.}\ \bibnamefont{Bruun}}\ and\ \bibinfo
  {author} {\bibfnamefont{E.}~\bibnamefont{Taylor}},\ }%
  \bibfield{journal}{%
  \bibinfo {journal} {Phys. Rev. Lett.}\ }%
  \textbf{\bibinfo {volume} {101}},\ \bibinfo {pages} {245301} (\bibinfo {year}
  {2008})%
  \bibAnnoteFile{NoStop}{BRU08}%
\bibitem{CON10}%
  \BibitemOpen
  \bibfield{author}{%
  \bibinfo {author} {\bibfnamefont{C.}~\bibnamefont{Wu}}\ and\ \bibinfo
  {author} {\bibfnamefont{J.~E.}\ \bibnamefont{Hirsch}},\ }%
  \bibfield{journal}{%
  \bibinfo {journal} {Phys. Rev. B}\ }%
  \textbf{\bibinfo {volume} {81}},\ \bibinfo {pages} {020508} (\bibinfo {year}
  {2010})%
  \bibAnnoteFile{NoStop}{CON10}%
\bibitem{ZHA10}%
  \BibitemOpen
  \bibfield{author}{%
  \bibinfo {author} {\bibfnamefont{C.}~\bibnamefont{Zhao}}, \bibinfo {author}
  {\bibfnamefont{L.}~\bibnamefont{Jiang}}, \bibinfo {author}
  {\bibfnamefont{X.}~\bibnamefont{Liu}}, \bibinfo {author}
  {\bibfnamefont{W.~M.}\ \bibnamefont{Liu}}, \bibinfo {author}
  {\bibfnamefont{X.}~\bibnamefont{Zou}},\ and\ \bibinfo {author}
  {\bibfnamefont{H.}~\bibnamefont{Pu}},\ }%
  \bibfield{journal}{%
  \bibinfo {journal} {Phys. Rev. A}\ }%
  \textbf{\bibinfo {volume} {81}},\ \bibinfo {pages} {063642} (\bibinfo {year}
  {2010})%
  \bibAnnoteFile{NoStop}{ZHA10}%
\bibitem{LIA10}%
  \BibitemOpen
  \bibfield{author}{%
  \bibinfo {author} {\bibfnamefont{R.}~\bibnamefont{Liao}}\ and\ \bibinfo
  {author} {\bibfnamefont{J.}~\bibnamefont{Brand}},\ }%
  \bibfield{journal}{%
  \bibinfo {journal} {Phys. Rev. A}\ }%
  \textbf{\bibinfo {volume} {82}},\ \bibinfo {pages} {063624} (\bibinfo {year}
  {2010})%
  \bibAnnoteFile{NoStop}{LIA10}%
\bibitem{MIY08}%
  \BibitemOpen
  \bibfield{author}{%
  \bibinfo {author} {\bibfnamefont{T.}~\bibnamefont{Miyakawa}}, \bibinfo
  {author} {\bibfnamefont{T.}~\bibnamefont{Sogo}},\ and\ \bibinfo {author}
  {\bibfnamefont{H.}~\bibnamefont{Pu}},\ }%
  \bibfield{journal}{%
  \bibinfo {journal} {Phys. Rev. A}\ }%
  \textbf{\bibinfo {volume} {77}},\ \bibinfo {pages} {061603} (\bibinfo {year}
  {2008})%
  \bibAnnoteFile{NoStop}{MIY08}%
\bibitem{SOG09}%
  \BibitemOpen
  \bibfield{author}{%
  \bibinfo {author} {\bibfnamefont{T.}~\bibnamefont{Sogo}}, \bibinfo {author}
  {\bibfnamefont{L.}~\bibnamefont{He}}, \bibinfo {author}
  {\bibfnamefont{T.}~\bibnamefont{Miyakawa}}, \bibinfo {author}
  {\bibfnamefont{S.}~\bibnamefont{Yi}}, \bibinfo {author}
  {\bibfnamefont{H.}~\bibnamefont{Lu}},\ and\ \bibinfo {author}
  {\bibfnamefont{H.}~\bibnamefont{Pu}},\ }%
  \bibfield{journal}{%
  \bibinfo {journal} {New J. Phys.}\ }%
  \textbf{\bibinfo {volume} {11}},\ \bibinfo {pages} {055017} (\bibinfo {year}
  {2009})%
  \bibAnnoteFile{NoStop}{SOG09}%
\bibitem{ZHA09}%
  \BibitemOpen
  \bibfield{author}{%
  \bibinfo {author} {\bibfnamefont{J.-N.}\ \bibnamefont{Zhang}}\ and\ \bibinfo
  {author} {\bibfnamefont{S.}~\bibnamefont{Yi}},\ }%
  \bibfield{journal}{%
  \bibinfo {journal} {Phys. Rev. A}\ }%
  \textbf{\bibinfo {volume} {80}},\ \bibinfo {pages} {053614} (\bibinfo {year}
  {2009})%
  \bibAnnoteFile{NoStop}{ZHA09}%
\bibitem{BLA10}%
  \BibitemOpen
  \bibfield{author}{%
  \bibinfo {author} {\bibfnamefont{D.}~\bibnamefont{Baillie}}\ and\ \bibinfo
  {author} {\bibfnamefont{P.~B.}\ \bibnamefont{Blakie}},\ }%
  \bibfield{journal}{%
  \bibinfo {journal} {Phys. Rev. A}\ }%
  \textbf{\bibinfo {volume} {82}},\ \bibinfo {pages} {033605} (\bibinfo {year}
  {2010})%
  \bibAnnoteFile{NoStop}{BLA10}%
\bibitem{STE02}%
  \BibitemOpen
  \bibfield{author}{%
  \bibinfo {author} {\bibfnamefont{S.}~\bibnamefont{Giovanazzi}}, \bibinfo
  {author} {\bibfnamefont{A.}~\bibnamefont{Gorlitz}},\ and\ \bibinfo {author}
  {\bibfnamefont{T.}~\bibnamefont{Pfau}},\ }%
  \bibfield{journal}{%
  \bibinfo {journal} {Phys. Rev. Lett.}\ }%
  \textbf{\bibinfo {volume} {89}},\ \bibinfo {pages} {130401} (\bibinfo {year}
  {2002})%
  \bibAnnoteFile{NoStop}{STE02}%
\bibitem{RON10}%
  \BibitemOpen
  \bibfield{author}{%
  \bibinfo {author} {\bibfnamefont{S.}~\bibnamefont{Ronen}}\ and\ \bibinfo
  {author} {\bibfnamefont{J.~L.}\ \bibnamefont{Bohn}},\ }%
  \bibfield{journal}{%
  \bibinfo {journal} {Phys. Rev. A}\ }%
  \textbf{\bibinfo {volume} {81}},\ \bibinfo {pages} {033601} (\bibinfo {year}
  {2010})%
  \bibAnnoteFile{NoStop}{RON10}%
\bibitem{CHA10}%
  \BibitemOpen
  \bibfield{author}{%
  \bibinfo {author} {\bibfnamefont{C.-K.}\ \bibnamefont{Chan}}, \bibinfo
  {author} {\bibfnamefont{C.}~\bibnamefont{Wu}}, \bibinfo {author}
  {\bibfnamefont{W.-C.}\ \bibnamefont{Lee}},\ and\ \bibinfo {author}
  {\bibfnamefont{S.}~\bibnamefont{DasSarma}},\ }%
  \bibfield{journal}{%
  \bibinfo {journal} {Phys. Rev. A}\ }%
  \textbf{\bibinfo {volume} {81}},\ \bibinfo {pages} {023602}%
  \bibAnnoteFile{NoStop}{CHA10}%
\bibitem{GOR03}%
  \BibitemOpen
  \bibfield{author}{%
  \bibinfo {author} {\bibfnamefont{K.}~\bibnamefont{G\'{o}ral}}, \bibinfo
  {author} {\bibfnamefont{M.}~\bibnamefont{Brewczyk}},\ and\ \bibinfo {author}
  {\bibfnamefont{K.}~\bibnamefont{Rzazewski}},\ }%
  \bibfield{journal}{%
  \bibinfo {journal} {Phys. Rev. A}\ }%
  \textbf{\bibinfo {volume} {67}},\ \bibinfo {pages} {025601} (\bibinfo {year}
  {2003})%
  \bibAnnoteFile{NoStop}{GOR03}%
\bibitem{HE08}%
  \BibitemOpen
  \bibfield{author}{%
  \bibinfo {author} {\bibfnamefont{L.}~\bibnamefont{He}}, \bibinfo {author}
  {\bibfnamefont{J.-N.}\ \bibnamefont{Zhang}}, \bibinfo {author}
  {\bibfnamefont{Y.}~\bibnamefont{Zhang}},\ and\ \bibinfo {author}
  {\bibfnamefont{S.}~\bibnamefont{Yi}},\ }%
  \bibfield{journal}{%
  \bibinfo {journal} {Phys. Rev. A}\ }%
  \textbf{\bibinfo {volume} {77}},\ \bibinfo {pages} {031605}%
  \bibAnnoteFile{NoStop}{HE08}%
\bibitem{LIM10}%
  \BibitemOpen
  \bibfield{author}{%
  \bibinfo {author} {\bibfnamefont{A.~R.~P.}\ \bibnamefont{Lima}}\ and\
  \bibinfo {author} {\bibfnamefont{A.}~\bibnamefont{Pelster}},\ }%
  \bibfield{journal}{%
  \bibinfo {journal} {Phys. Rev. A}\ }%
  \textbf{\bibinfo {volume} {81}},\ \bibinfo {pages} {021606} (\bibinfo {year}
  {2010})%
  \bibAnnoteFile{NoStop}{LIM10}%
\bibitem{END10}%
  \BibitemOpen
  \bibfield{author}{%
  \bibinfo {author} {\bibfnamefont{Y.}~\bibnamefont{Endo}}, \bibinfo {author}
  {\bibfnamefont{T.}~\bibnamefont{Miyakawa}},\ and\ \bibinfo {author}
  {\bibfnamefont{T.}~\bibnamefont{Nikuni}},\ }%
  \bibfield{journal}{%
  \bibinfo {journal} {Phys. Rev. A}\ }%
  \textbf{\bibinfo {volume} {81}},\ \bibinfo {pages} {063624} (\bibinfo {year}
  {2010})%
  \bibAnnoteFile{NoStop}{END10}%
\bibitem{ZYI10}%
  \BibitemOpen
  \bibfield{author}{%
  \bibinfo {author} {\bibfnamefont{J.-N.}\ \bibnamefont{Zhang}}\ and\ \bibinfo
  {author} {\bibfnamefont{S.}~\bibnamefont{Yi}},\ }%
  \bibfield{journal}{%
  \bibinfo {journal} {Phys. Rev. A}\ }%
  \textbf{\bibinfo {volume} {81}},\ \bibinfo {pages} {033617} (\bibinfo {year}
  {2010})%
  \bibAnnoteFile{NoStop}{ZYI10}%
\bibitem{SIM06}%
  \BibitemOpen
  \bibfield{author}{%
  \bibinfo {author} {\bibfnamefont{A.}~\bibnamefont{Altland}}\ and\ \bibinfo
  {author} {\bibfnamefont{B.}~\bibnamefont{Simons}},\ }%
  \emph{\bibinfo {title} {Condensed Matter Field Theory}}\ (\bibinfo
  {publisher} {Cambridge University Press},\ \bibinfo {address} {Cambridge},\
  \bibinfo {year} {2006})\ \bibinfo {note} {(page 250)}%
  \bibAnnoteFile{NoStop}{SIM06}%
\bibitem{VOL90}%
  \BibitemOpen
  \bibfield{author}{%
  \bibinfo {author} {\bibfnamefont{D.}~\bibnamefont{Vollhardt}}\ and\ \bibinfo
  {author} {\bibfnamefont{P.}~\bibnamefont{Wolfle}},\ }%
  \emph{\bibinfo {title} {The superfluid phases of Helium 3}}\ (\bibinfo
  {publisher} {Taylor and Francis},\ \bibinfo {address} {Landon},\ \bibinfo
  {year} {1990})%
  \bibAnnoteFile{NoStop}{VOL90}%
\bibitem{GUR07}%
  \BibitemOpen
  \bibfield{author}{%
  \bibinfo {author} {\bibfnamefont{V.}~\bibnamefont{Gurarie}}\ and\ \bibinfo
  {author} {\bibfnamefont{L.}~\bibnamefont{Radzihovsky}},\ }%
  \bibfield{journal}{%
  \bibinfo {journal} {Ann. Phys.}\ }%
  \textbf{\bibinfo {volume} {322}},\ \bibinfo {pages} {2} (\bibinfo {year}
  {2007})%
  \bibAnnoteFile{NoStop}{GUR07}%
\bibitem{PAR07}%
  \BibitemOpen
  \bibfield{author}{%
  \bibinfo {author} {\bibfnamefont{M.~M.}\ \bibnamefont{Parish}}, \bibinfo
  {author} {\bibfnamefont{F.~M.}\ \bibnamefont{Marchetti}}, \bibinfo {author}
  {\bibfnamefont{A.}~\bibnamefont{Lamacraft}},\ and\ \bibinfo {author}
  {\bibfnamefont{B.~D.}\ \bibnamefont{Simons}},\ }%
  \bibfield{journal}{%
  \bibinfo {journal} {Phys. Rev. Lett.}\ }%
  \textbf{\bibinfo {volume} {98}},\ \bibinfo {pages} {160402} (\bibinfo {year}
  {2007})%
  \bibAnnoteFile{NoStop}{PAR07}%
\bibitem{ZHU06}%
  \BibitemOpen
  \bibfield{author}{%
  \bibinfo {author} {\bibfnamefont{L.}~\bibnamefont{He}}, \bibinfo {author}
  {\bibfnamefont{M.}~\bibnamefont{Jin}},\ and\ \bibinfo {author}
  {\bibfnamefont{P.}~\bibnamefont{Zhuang}},\ }%
  \bibfield{journal}{%
  \bibinfo {journal} {Phys. Rev. B}\ }%
  \textbf{\bibinfo {volume} {74}},\ \bibinfo {pages} {024516} (\bibinfo {year}
  {2006})%
  \bibAnnoteFile{NoStop}{ZHU06}%
\bibitem{GRI06}%
  \BibitemOpen
  \bibfield{author}{%
  \bibinfo {author} {\bibfnamefont{E.}~\bibnamefont{Taylor}}, \bibinfo {author}
  {\bibfnamefont{A.}~\bibnamefont{Griffin}}, \bibinfo {author}
  {\bibfnamefont{N.}~\bibnamefont{Fukushima}},\ and\ \bibinfo {author}
  {\bibfnamefont{Y.}~\bibnamefont{Ohashi}},\ }%
  \bibfield{journal}{%
  \bibinfo {journal} {Phys. Rev. A}\ }%
  \textbf{\bibinfo {volume} {74}},\ \bibinfo {pages} {063626} (\bibinfo {year}
  {2006})%
  \bibAnnoteFile{NoStop}{GRI06}%
\bibitem{RAD03}%
  \BibitemOpen
  \bibfield{author}{%
  \bibinfo {author} {\bibfnamefont{H.~A.}\ \bibnamefont{Radovan}}, \bibinfo
  {author} {\bibfnamefont{N.~A.}\ \bibnamefont{Fortune}}, \bibinfo {author}
  {\bibfnamefont{T.~P.}\ \bibnamefont{Murphy}}, \bibinfo {author}
  {\bibfnamefont{S.~T.}\ \bibnamefont{Hannahs}}, \bibinfo {author}
  {\bibfnamefont{E.~C.}\ \bibnamefont{Palm}}, \bibinfo {author}
  {\bibfnamefont{S.~W.}\ \bibnamefont{Tozer}},\ and\ \bibinfo {author}
  {\bibfnamefont{D.}~\bibnamefont{Hall}},\ }%
  \bibfield{journal}{%
  \bibinfo {journal} {Nature}\ }%
  \textbf{\bibinfo {volume} {425}},\ \bibinfo {pages} {51} (\bibinfo {year}
  {2003})%
  \bibAnnoteFile{NoStop}{RAD03}%
\bibitem{LIAO10}%
  \BibitemOpen
  \bibfield{author}{%
  \bibinfo {author} {\bibfnamefont{Y.-a.}\ \bibnamefont{Liao}}, \bibinfo
  {author} {\bibfnamefont{A.~S.~C.}\ \bibnamefont{Rittner}}, \bibinfo {author}
  {\bibfnamefont{T.}~\bibnamefont{Paprotta}}, \bibinfo {author}
  {\bibfnamefont{W.}~\bibnamefont{Li}}, \bibinfo {author}
  {\bibfnamefont{G.~B.}\ \bibnamefont{Patridge}}, \bibinfo {author}
  {\bibfnamefont{R.~G.}\ \bibnamefont{Hulet}}, \bibinfo {author}
  {\bibfnamefont{S.~K.}\ \bibnamefont{Baur}},\ and\ \bibinfo {author}
  {\bibfnamefont{E.~J.}\ \bibnamefont{Mueller}},\ }%
  \bibfield{journal}{%
  \bibinfo {journal} {Nature}\ }%
  \textbf{\bibinfo {volume} {467}},\ \bibinfo {pages} {567} (\bibinfo {year}
  {2010})%
  \bibAnnoteFile{NoStop}{LIAO10}%
\bibitem{REG05}%
  \BibitemOpen
  \bibfield{author}{%
  \bibinfo {author} {\bibfnamefont{C.~A.}\ \bibnamefont{Regal}}, \bibinfo
  {author} {\bibfnamefont{M.}~\bibnamefont{Greiner}}, \bibinfo {author}
  {\bibfnamefont{S.}~\bibnamefont{Giorgini}}, \bibinfo {author}
  {\bibfnamefont{M.}~\bibnamefont{Holland}},\ and\ \bibinfo {author}
  {\bibfnamefont{D.~S.}\ \bibnamefont{Jin}},\ }%
  \bibfield{journal}{%
  \bibinfo {journal} {Phys. Rev. Lett.}\ }%
  \textbf{\bibinfo {volume} {95}},\ \bibinfo {pages} {250404} (\bibinfo {year}
  {2005})%
  \bibAnnoteFile{NoStop}{REG05}%
\bibitem{CHI04}%
  \BibitemOpen
  \bibfield{author}{%
  \bibinfo {author} {\bibfnamefont{C.}~\bibnamefont{Chin}}, \bibinfo {author}
  {\bibfnamefont{M.}~\bibnamefont{Bartenstein}}, \bibinfo {author}
  {\bibfnamefont{A.}~\bibnamefont{Altmeyer}}, \bibinfo {author}
  {\bibfnamefont{S.}~\bibnamefont{Riedl}}, \bibinfo {author}
  {\bibfnamefont{S.}~\bibnamefont{Jochim}}, \bibinfo {author}
  {\bibfnamefont{J.}~\bibnamefont{{Hecker Denschlag}}},\ and\ \bibinfo {author}
  {\bibfnamefont{R.}~\bibnamefont{Grimm}},\ }%
  \bibfield{journal}{%
  \bibinfo {journal} {Science}\ }%
  \textbf{\bibinfo {volume} {305}},\ \bibinfo {pages} {1128} (\bibinfo {year}
  {2004})%
  \bibAnnoteFile{NoStop}{CHI04}%
\bibitem{GAE10}%
  \BibitemOpen
  \bibfield{author}{%
  \bibinfo {author} {\bibfnamefont{J.}~\bibnamefont{Gaebler}}, \bibinfo
  {author} {\bibfnamefont{J.}~\bibnamefont{Stewart}}, \bibinfo {author}
  {\bibfnamefont{T.}~\bibnamefont{Drake}}, \bibinfo {author}
  {\bibfnamefont{D.}~\bibnamefont{Jin}}, \bibinfo {author}
  {\bibfnamefont{A.}~\bibnamefont{Perali}}, \bibinfo {author}
  {\bibfnamefont{P.}~\bibnamefont{Pieri}},\ and\ \bibinfo {author}
  {\bibfnamefont{G.}~\bibnamefont{Strinati}},\ }%
  \bibfield{journal}{%
  \bibinfo {journal} {Nat. Phys.}\ }%
  \textbf{\bibinfo {volume} {6}},\ \bibinfo {pages} {569} (\bibinfo {year}
  {2010})%
  \bibAnnoteFile{NoStop}{GAE10}%
\end{thebibliography}
\end{document}